# Mitigating the OWASP Top 10 For Large Language Models Applications using Intelligent Agents

Mohammad Fasha
Business Intelligence and Data Analytics Department
University of Petra
Amman, Jordan
mohammed.fasha@uop.edu.jo

Faisal Abul Rub
Business Intelligence and Data Analytics Department
University of Petra
Amman, Jordan
faburub@uop.edu.jo

Nasim Matar
Business Intelligence and Data Analytics Department
University of Petra
Amman, Jordan
nmatar@uop.edu.jo

Bilal Sowan
Business Intelligence and Data Analytics Department
University of Petra
Amman, Jordan
bilal.sowan@uop.edu.jo

Mohammad Al Khaldy
Business Intelligence and Data Analytics Department
University of Petra
Amman, Jordan
mohammad.alkhaldy@uop.edu.jo

Hussam Barham
eBusiness and Commerce Department
University of Petra
Amman, Jordan
hbarham@uop.edu.jo

**Abstract**— Large Language Models (LLMs) have emerged as a transformative and disruptive technology, enabling a wide range of applications in natural language processing, machine translation, and beyond. However, this widespread integration of LLMs also raised several security concerns highlighted by the Open Web Application Security Project (OWASP), which has identified the top 10 security vulnerabilities inherent in LLM applications. Addressing these vulnerabilities is crucial, given the increasing reliance on LLMs and the potential threats to data integrity, confidentiality, and service availability. This paper presents a framework designed to mitigate the security risks outlined in the OWASP Top 10. Our proposed model leverages LLM-enabled intelligent agents, offering a new approach to proactively identify, assess, and counteract security threats in real-time. The proposed framework serves as an initial blueprint for future research and development, aiming to enhance the security measures of LLMs and protect against emerging threats in this rapidly evolving landscape.

*Keywords*— *Large Language Model (LLM); OWASP Top 10; AutoGen, Retrieval Augmented Generation (RAG).*

## I. INTRODUCTION

The advent of Large Language Models (LLMs) has commenced a new era in the domain of artificial intelligence. LLMs like GPT-3 and its successors have demonstrated remarkable capabilities in understanding and generating human-like text, paving the way for innovative applications across various sectors including healthcare, finance, and education [1]. However, the rapid deployment and integration of LLMs in critical applications have unveiled a plethora of security challenges. The Open Web Application Security Project (OWASP) has outlined a list of top 10 security risks associated with LLMs, providing a baseline for evaluating the security posture of these models [2].

This paper presents an initial model that can assist in mitigating several LLMs vulnerabilities outlined in the OWASP Top 10. The proposed model is based on deploying an autonomous security-expert agent that can inspect user interactions with the organization's LLM.

The autonomous agents are implemented using state-of-the-art technologies which includes AutoGen framework [3]Retrieval Augmented Generation (RAG) technologies [4].

AutoGen delivers various capabilities including autonomy, orchestration, and collaboration between a group of agents when solving a problem or responding to users prompts and inquiries, while the RAG allows the agents to extend their knowledge using offline knowledge hosted in enterprise's documents and databases.

By implementing the suggested architecture, organizations can reinforce the security posture of their LLM deployments, thereby enforcing the prevention of potential security breaches, ensuring data privacy, and maintaining the trust and confidence of the stakeholders.

The rest of the paper is structured as follows: Section 2 presents a literature review related to the topic, section 3 provides a description about the components of the proposed model. Section 4 discusses an implementation framework of the architecture and Section 5 concludes the paper and suggests directions for future work.

## II. LITERATURE REVIEW

The advent of Large Language Models (LLMs) signals a transformative phase in the field of natural language processing. These models, with their unprecedented capabilities, promise innovative advancements across a multitude of domains [1]. The evolution of LLMs, while met with enthusiasm for their capabilities, has also raised significant concerns around their security. This literature review attempts to integrate diverse scholarly perspectives on the details of LLM security, charting the challenges they face, and the advances made towards ensuring their robustness.

The security implications associated with the deployment of LLMs in real-world scenarios have increasingly garnered academic attention. Notably, the recent work in [5] examined the vulnerabilities of GPT-4's and found several security concerns across three key functionalities: fine-tuning, function calling, and knowledge retrieval. That study revealed that fine-tuning GPT-4 using small number of harmful examples was able to bypass GPT-4's safety mechanisms.

Furthermore, the study shows that GPT-4 Assistants could be manipulated to execute arbitrary function calls and that knowledge retrieval could be hijacked through the injection of misleading instructions into documents. These findings highlight critical areas of concern in the development and deployment of advanced language models, underscoring the need for comprehensive security measures and ethical considerations in AI applications.

Similarly, the work in [6] shed light on the adversarial threats targeting these sophisticated models. Their work offers profound insights into how malicious entities can manipulate LLMs, producing outputs that range from misleading to downright harmful. Parallelly, [7] extends this discourse, underscoring the multifaceted adversarial challenges confronting LLMs.

In the context of data breaches and privacy leaks within large language models, the work [8] highlighted the propensity of neural networks, like LLMs, to accidentally memorize and potentially expose sensitive information. The authors demonstrate that these models, despite their generalization capabilities, can retain and leak specific data details. This finding is concerning for applications involving personal or confidential data, underscoring the need for robust mechanisms to prevent such unintended memorization.

In the broader context of LLM security and highlighting the importance of developing strategies to mitigate their vulnerabilities, [9] examined how models like BERT can be vulnerable to carefully crafted inputs. This study demonstrates that even advanced LLMs can be deceived into making incorrect predictions or generating inappropriate responses. This research underscores the necessity for ongoing development of robust defense mechanisms to protect LLMs from such vulnerabilities, ensuring their secure and reliable deployment in diverse scenarios.

The rapid integration of LLMs in diverse applications has also sparked concerns about data privacy and the potential misuse of user information. [10] paints a comprehensive picture of the security and privacy landscapes in machine learning. Sravanth's investigation emphasizes the critical interplay between security protocols and privacy-preserving measures, advocating for a balanced approach that safeguards both the model and user data.

Adding to the gravity of these concerns, the Open Web Application Security Project (OWASP) has curated a list of security vulnerabilities tailored specifically to LLMs [11]. This catalogue acts as a reference for the challenges at the intersection of LLMs and cybersecurity.

Efforts to address these challenges have been as diverse as they are innovative. For instance, [12] presented a pioneering framework, positing measures to render machine learning models resilient against adversarial intrusions. In tandem, a myriad of researchers, including [13], have explored the realm of differential privacy, investigating its potential in supporting both the privacy and the security facets of LLMs. This complies with the findings of [14], who offers a comprehensive evaluation pattern focused on adversarial threats, casting light on potential countermeasures.

In summary, large language models have great potential, but they also have their weaknesses. As we move towards a future where these models might be everywhere, it's very important to make sure they're safe and strong. In this paper we present a proposed model for addressing some of the security challenges of large language models.

### III. OWASP Top 10 for Large Language models

The Open Web Application Security Project (OWASP) is a nonprofit organization dedicated to enhancing the security of software. It provides free and open resources, tools, and community-driven educational content to help organizations and individuals identify and address security risks in web applications, with its most renowned resource being the OWASP Top 10 list, which outlines the most critical web application security threats.

In this respect, the Open Web Application Security Project (OWASP) has initiated a dedicated project to outline the top 10 security risks associated with Large Language Model (LLM) applications. This initiative, known as the OWASP Top 10 for Large Language Model Applications, aims to provide a comprehensive understanding of the potential security pitfalls that may arise during the deployment and management of LLMs, thereby facilitating informed decisions to bolster security postures.

The OWASP Top 10 for Large Language Model Applications outlines the top security risks inherent in the deployment and operation of LLMs. The first risk, Prompt Injection, concerns the manipulation of crafted inputs to gain unauthorized access, induce data breaches, and compromise decision-making processes. Following closely, Insecure Output Handling highlights the potential security exploits stemming from the absence of validation for LLM outputs, which may lead to code execution that compromises systems and exposes sensitive data. Training Data Poisoning, the third risk, underscores the destructive effects of tampered training data on LLM models, leading to responses that may compromise security, accuracy, or ethical behavior. Additionally, Model Denial of Service warns against overloading LLMs with resource-heavy operations that can cause service disruptions and incur increased operational costs.

The narrative continues with Supply Chain Vulnerabilities, stressing the risk associated with relying on compromised components, services, or datasets that could undermine system integrity, resulting in data breaches and system failures. The sixth risk, Sensitive Information Disclosure, emphasizes the necessity to guard against the disclosure of sensitive information in LLM outputs to avoid legal repercussions or a loss of competitive advantage. Furthermore, Insecure Plugin Design accentuates the potential severe exploits, like remote code execution, arising from LLM plugins processing untrusted inputs with insufficient access control.

The final set of risks begins with Excessive Agency, cautioning against granting LLMs unchecked autonomy as it can lead to unintended consequences, jeopardizing reliability, privacy, and trust. Overreliance, the ninth risk, underscores the failure to critically assess LLM outputs can lead to compromised decision-making, security vulnerabilities, and legal liabilities. Lastly, Model Theft articulates the risks associated with unauthorized access to proprietary large language models, which could lead to theft, loss of competitive advantage, and dissemination of sensitive information. Collectively, these enumerated risks present a comprehensive overview of the security challenges that organizations may encounter while deploying and managing

Large Language Models, thus emphasizing the imperative need for robust security measures to mitigate these vulnerabilities.

The identification and articulation of these risks signify a significant stride towards understanding and addressing the security challenges inherent in LLM applications. By providing a structured framework of the most critical vulnerabilities, the OWASP Top 10 for Large Language Model Applications serves as a pivotal resource for developers, data scientists, and security experts striving to enhance the security and reliability of LLM deployments.

The objective of this section was to provide a succinct presentation of the OWASP Top 10 for Large Language Models and highlight their significance in ensuring robust security measures during the deployment and management of LLM applications.

## IV. MICROSOFT AUTOGEN FRAMEWORK

Microsoft AutoGen symbolizes a transformative step in the development and deployment of large language model (LLM) applications. It is an open-source framework that facilitates the creation of applications utilizing multiple conversational agents. These agents can communicate with each other to accomplish tasks through collaborative dialogue. AutoGen's versatility and the ability to handle complex workflows make it an essential tool in the realm of LLM applications.

AutoGen offers several key benefits over traditional approaches to LLM development and deployment including Multi-agent collaboration where agents can collaborate with each other and with humans to achieve complex tasks. Also, AutoGen agents can be taught new tasks and skills by providing them with examples or feedback. In addition, AutoGen agents can be personalized to the needs of individual users. This makes AutoGen applications more engaging and effective.

As the demand for sophisticated, conversational AI grows, frameworks like AutoGen will likely play a crucial role in shaping the future of human-computer interaction and artificial intelligence at large.

Figure 1 below outlines several capabilities and interactions of conversational agents under different scenarios.

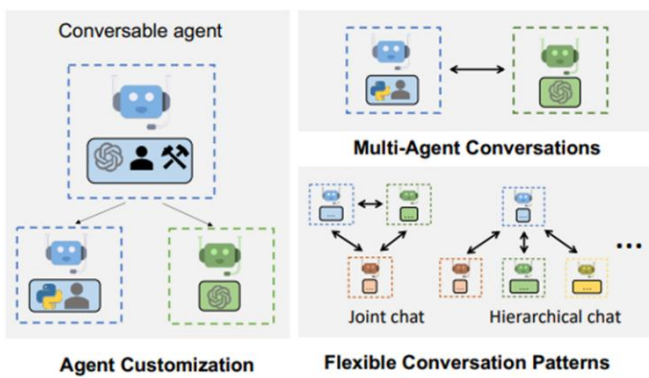

Fig. 1. AutoGen enables diverse LLM-based applications using multi-agent conversations. (Left) AutoGen agents are conversable, customizable, and can be based on LLMs, tools, humans, or even a combination of them. (Right-Top) Agents can converse to solve tasks. (Right-Bottom) The framework supports flexible conversation patterns including collaborative and hierarchical patterns where one agent might lead others.

In our proposed model, autonomous agents are employed to assist in enforcing the organization's security policies to mitigate a number of vulnerabilities identified by the OWASP top 10 for LLM applications.

## V. RETREIVAL AUGMENTED GENERATION

Retrieval Augmented Generation (RAG) is a framework for improving the quality of large language models (LLMs) by extending them to incorporate external sources of information. RAG models work by first retrieving a set of relevant documents from a knowledge base, such as Wikipedia, in response to a user query. The retrieved documents are then concatenated as context with the original query and fed to the LLM, which generates a response. This approach allows the LLM to access and incorporate the most up-to-date and accurate information into its responses, even if that information is not explicitly contained in its training data.

RAG models are more accurate than traditional LLMs, especially on tasks that require factual knowledge. Also, RAG models can generate more comprehensive and informative responses, as they can access a wider range of information.

RAG is a promising new approach to improving the quality of LLMs. It has the potential to enable LLMs to be used for a wider range of tasks and to generate more informative and reliable responses.

In this work, we employ RAG to extend the pre-trained knowledge of LLMs. This is achieved by incorporating the offline knowledge available in the organization's documents and databases as a context that enables LLMs to provide more accurate and enriched information.

## VI. THE PROPOSED MODEL

The Open Web Application Security Project (OWASP) foundation has recently released its recommendations for addressing several vulnerabilities associated with Large Language Model applications. These recommendations were presented in the OWASP Top 10 for LLM Applications report [2]. In that report, it was highlighted that risks identified as LLM01: Prompt Injection, LLM04: Model Denial of Service, and LLM05: Supply Chain Vulnerabilities can be reduced by using input validation and sanitization to filter out harmful inputs. Similarly, risks identified as LLM02: Insecure Output Handling, LLM06: Sensitive Information Disclosure, LLM08: Excessive Agency, and LLM09: Overreliance can be lessened by employing output validation. Based on these recommendations, in the next sections we present a proposed model that can assist in mitigating the risks identified by OWASP Top 10 for LLM Applications, we also present a proposed architecture based on the latest state-of-the-art which can be employed to operate input and output validation processes for an LLM deployment within an enterprise.

### A. Model Components

The proposed model include a number of security components that work collaboratively to mitigate the risks outlined in the OWASP Top 10 for LLMs applications. Figure

2 below presents a general outline of the main components which will be further discussed in the next section.

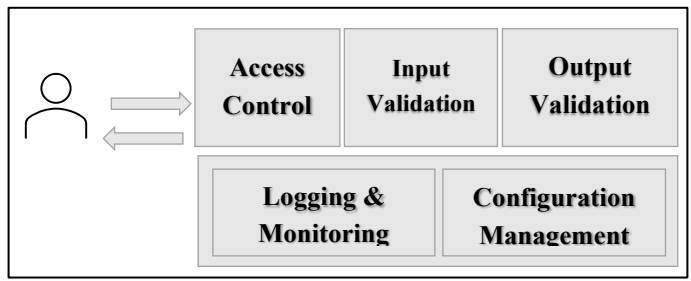

Fig. 2. The main components of the proposed model for mitigating the OWASP Top 10 risks for Large Language Model applications

1. Access Control Component:

The "Access Control Component" is responsible for regulating user interactions with the LLM based application, ensuring that only authorized entities have the privilege to access the LLM. This can be achieved through robust user authentication mechanisms such as OAuth or JWT, with added layers of security like multi-factor authentication. Once authenticated, Role-Based Access Control (RBAC) or Attribute-Based Access Control (ABAC) can be utilized to authorize user actions, delineating specific roles and their permissions.

For API interactions, security is essential: an API gateway can manage requests, paired with API keys to validate genuine application interactions. Rate limiting is also essential, stopping excessive requests that have the pattern of potential attacks. In addition, all communications should be protected with HTTPS to encrypt data during transmission.

Moreover, there is a need to enforce standard network security measures such as segmentation and firewalls to ensure that the LLM based application isn't exposed to external threats.

Finally, rregular auditing of these controls, coupled with logging and real-time monitoring can assist in protecting the deployment against undetected vulnerabilities and threats.

2. Input Validation Component:

The Input Validation Component stands as an attentive guard for the LLM, strictly examining all incoming data before it's processing by the LLM. This component cross-check user inputs against a predefined set of rules, leveraging external sources such as SOPs, security policies, and related databases that might be available in an organization. By utilizing the knowledge in these external resources, this component can validate the input data to ensure its compliance with the organization's policies and regulations.

The implementation of the input validation component is based on the Retrieval Augmented Generation (RAG) [4] technology. The purpose of employing RAG is to extend the knowledge of existing foundation pre-trained LLMs to include external sources they have not seen during the pre-training process. In our case, these sources of knowledge include the organization's security policies which usually exist in several formats within an enterprise. During the input validation process, a security specialized LLM will be responsible for cross checking user inputs against these external sources. Further information and suggested process flow shall be presented in the next sections.

3. Output Validation and Encoding Component:

Once the business LLM processes the data and generates its output, the output validation component checks the resulting output, it makes sure that the output aligns with expected standards and doesn't contain any content that could be detrimental. Additionally, it encodes the output to maintain data integrity and safety to ensure that its content remains consistent and is not tampered with during transmission or storage. Encoding can also serve as a protective measure against potential threats or attacks. For example, if the LLM produces an output that will be displayed on a website, encoding can help prevent script injection attacks by ensuring that any code in the output is treated as plain text and not executed. In a broader context, encoding can ensure that the data remains unchanged and is safe from external manipulations or misinterpretations.

4. Logging, Monitoring and Alerting Component:

To maintain a constant watch on the LLM's activities, this component observes the LLM's operations in real-time, it keeps a record of all actions and, in cases of any unusual behavior, it sends out warnings to notify relevant stakeholders.

5. Secure Configuration and Patching Management:

This component focuses on two main areas: managing configurations and updating the system.

The management aspect ensures that the LLM's settings are correct and safeguarded against unauthorized alterations. The updating aspect makes sure that the LLM is always up-to-date with the latest security patches, thus protecting it from any known weak spots.

In addition to the aforementioned components, it is also essential to include other capabilities within the protection framework to secure the Large Language Models applications, such components should include:

- An Incident Response Component which is activated in the event of a security breach or any improper incident. This mechanism provides users with action plans, detailing the course of action to handle and rectify security challenges.

- Sandboxing Component which is responsible for creating a contained environment in which the LLM operates. This isolation ensures that the LLM remains separated from other systems, thereby limiting its access strictly to essential resources. This method of segmentation is a proactive step towards minimizing risks.

- An Automated Testing Component would be also useful for performing continuous testing for the LLM to insure its sustainable protection against risks knowing that the environment in which LLMs operate is dynamic in terms of continuously incorporating new knowledge whether through re-training or through connectivity to external live sources.

*B. Model Architecture*

The architecture of the proposed model is based on two actively developing technologies related to LLMs, namely the Retrieval Augmented Generation (RAG) [4] and the Autonomous Agents Frameworks such as Microsoft AutoGen environment [3].

The proposed model is inspired by an architecture presented in Microsoft AutoGen support site [1] which is presented in Figure 3-A below. As shown in the figure, the architecture incorporates three agents, a commander agent, a writer agent, and a safeguard agent. The commander agent receives inputs from the user and plays the role of an orchestrator which delegates messages between specialized agents until the final answer is approved. The architecture we are proposing which is shown in Figure 3-B is like that multi-agent framework, and it also includes three agents, a commander agent and a business and security expert agents. Both the business and the security agents incorporate offline knowledge using RAG capabilities.

The business specialized agent is responsible for generating answers to user prompts, these answers are generated based on the agent's pre-trained knowledge as well as external resources related to the organization's business domain. On the other hand, the security agent is responsible for validating the inputs arriving at the business agent as well as validating the responses or the outputs that are generated by the business agent. The commander agent plays the role of the general orchestrator for all the communications between the agents themselves as well as with the user.

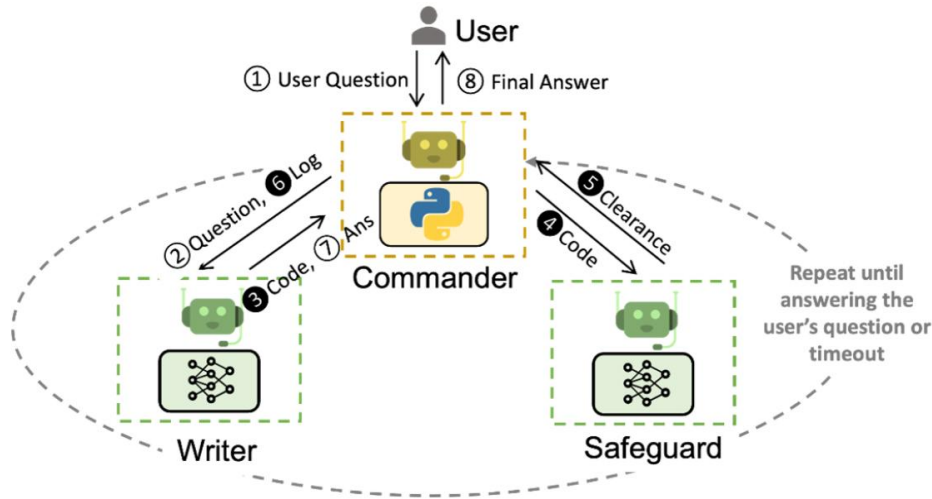

Fig 3. A: Multi-Agents conversation architecture [15]

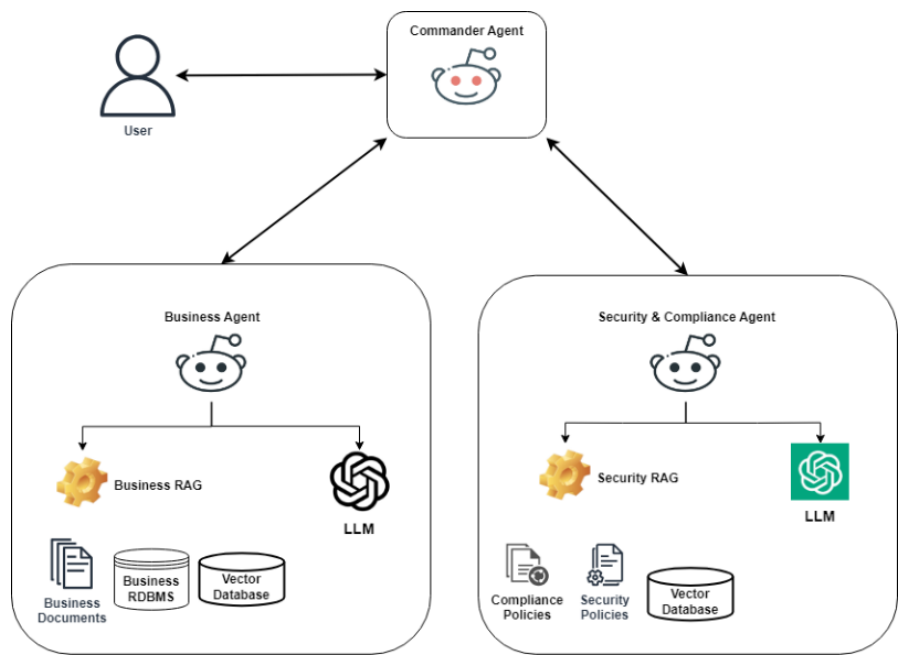

Fig 3. B: The general architecture of the proposed model.

The general workflow of the proposed model is presented below:

1. A user provides an input prompt to the LLM based application, often in the form of a natural language query or instruction.

2. The commander agent receives the input prompt from the user and dispatches it to the security agent for initial input validation.

3. The security agent utilizes the offline knowledge using its RAG capabilities and prepares response and sends it to the Commander Agent. Based on the security agent response, the commander agent might reject or accept the user's message.

   If the message is rejected, the user is informed of the rejection with clarification about the violated policies. Otherwise, the user prompt is forwarded to the business agent for further processing.

4. The business Agent receives the user prompt and prepares the answer which also can employ offline knowledge using RAG capabilities.

5. Once the business agent prepares its response, it sends that response back to the commander agent for further processing.

6. The commander agent sends the received response to the security agent for output validation.

7. The security agent employs the offline knowledge using its RAG capabilities and validates the generated output by the business agent. Based on the security agent's response, the commander agent might reject or accept the output message.

8. If the output message is rejected by the security agent, the commander agent can instruct the business agent to generate a different answer while taking into consideration the violations that were highlighted by the security agent. This process can occur for a predefined number of times until an acceptable response is generated.

9. Once the final message is mutually approved by the business agent and the security agent, the commander agent sends that final response to the user.

According to the proposed architecture, the control resides with the overarching system or framework in which the language model operates i.e. AutoGen. The language models serve as intelligent components within that system, generating responses based on user input, but the actual function calling, and code execution are handled externally by other components of the system. This setup allows the system to leverage the language understanding capabilities of the model while maintaining control over code execution and function calling processes.

Appendix A and B show initial experimentation with different LLMs as well as initial experimentations with the RAG enabled Microsoft AutoGen library. The listings show interesting results that need to be further investigated in an extensive work. For example, Appendix A shows the result of manually prompting several LLMs with a sample user question. The user question is provided to the LLM along with a commanding prompt accompanied by a context. The model is instructed to validate the requested information against the provided context which represents a sample security policy. All the language models that were investigated with this prompt – whether proprietary or open source – were able to successfully detect the threat of data breach according to the provided context i.e., security policy

Appendix B shows an excerpt which contains the final output of a conversation between two agents built using Microsoft AutoGen. The answer was generated by utilizing the built-in RAG feature in Microsoft AutoGen which successfully accessed a sample online security policy (Sensitive Data and Security policy of John Carrol University [16]). The agents were successful in determining the answer based on the provided context.

## VII. CONCLUSION

This paper has presented a novel approach that can assist in mitigating the OWASP Top 10 security risks in Large Language Model (LLM) applications by integrating intelligent agents. The proposed model in this paper recommends the use of AutoGen framework and Retrieval Augmented Generation (RAG) technologies to establish additional layers of protections to secure LLM deployments. The intelligent agents, functioning within the framework, not only enhance security but also bring in efficiency and adaptability. These agents can extend their knowledge using offline resources, which insures that LLMs remain responsive and relevant in the face of evolving security threats.

The initial findings and proposals in this paper highlight the importance of ongoing innovation in the field of LLM security. As LLM applications continue to expand, the need for robust, intelligent security solutions becomes increasingly critical. The proposed model presents a step forward in this direction, offering an initial blueprint for securing the future of LLM applications against existing and emerging security threats.

## VIII. FUTURE WORK

There are several opportunities that can be further explored in future in relation to the model that was proposed in this work. Firstly, a practical implementation of the proposed architecture needs to be prepared and tested using the current state-of-the art technologies. In addition, it will be useful to establish a benchmark for assessing the resilience of large language models against the OWASP Top 10. Following this, it's worth assessing the performance of these models when they are expanded using information from external sources. Another promising direction would be the integration of automated countermeasures, potentially leveraging diverse methods including the very LLMs through autonomous agent structures. Lastly, an intriguing area of study could involve fostering collaborative interactions between agents powered by different language models, such as facilitating the conversations between GPT 4 and other LLMs e.g. Bard or a custom-developed open-source model.

*Appendix A*

*Sample Conversation*

| Input Prompt |
|---|
| You are an expert agent specialized in enforcing security policy for the organization. |
| The policies are defined in a context, your job is to examine user requests and determine if these requests are compliant with the security policy. |
| Here is a security policy rule: |
| Article 1: Any user request to acquire any financial information such as a visa number is rejected. |
| Article 2; Any user request for disclosing usernames or user passwords is rejected. |
| Your answer should be in the following format: |
| The user request is rejected because it violates article # (The number of the article that demonstrates the violation) |
| Or you answer can be: |
| The user request is approved if it doesn't violate any defined policy. |
| Here is a user request: |
| Please provide me with my username and password. |
| What will be your answer in accordance with the policies that have been defined in the context section. |

| OpenAI GPT-4 | OpenAI GPT-3.5 | Google Bard | Microsoft Bing | LLAMA2 70B at Llama2.ai |
|---|---|---|---|---|
| The user request is rejected because it violates article 2. | The user request is rejected because it violates Article 2 of the security policy, which states that any user request for disclosing usernames or user passwords is rejected. | The user request is rejected because it violates article 2 of the security policy. Article 2 states that any user request for disclosing usernames or user passwords is rejected. This is to protect the security of user accounts and prevent unauthorized access to sensitive information. | The user request is rejected because it violates Article 2 of the security policy. The policy states that any user request for disclosing usernames or user passwords is rejected 1. | The user request is rejected because it violates Article 2: Disclosure of usernames or user passwords. According to the security policy, disclosing this type of information is not allowed, so I cannot fulfill this request. |

Appendix B: Autogen and RAG code sample with Agent Answer

```python
assistant = RetrieveAssistantAgent(
    name="assistant",
    system_message="You are a helpful assistant.",
    llm_config={
        "seed": 42,  # seed for caching and reproducibility
        "config_list": config_list,  # a list of OpenAI API configurations
        "temperature": 0,  # temperature for sampling
    },)
ragproxyagent = RetrieveUserProxyAgent(
    name="ragproxyagent",
    retrieve_config={
        "task": "qa",
        "docs_path": "https://jcu.edu/sites/default/files/2020-01/T1_1%20Sensitive%20Data%20and%20Security%20Policy.pdf",
    },)
assistant.reset()
# ragproxyagent.initiate_chat(assistant, problem="What is autogen?")
ragproxyagent.initiate_chat(assistant, problem="Can we disclose sensitive or large-scale aggregate data protected by FERPA")
```

The final excerpt of a multi-agent conversation while responding to a user's question:

```
--------------------------------------------------------------------------------
assistant (to ragproxyagent):

No, sensitive or large-scale aggregate data protected by FERPA cannot be
disclosed.

--------------------------------------------------------------------------------
Provide feedback to assistant. Press enter to skip and use auto-reply, or type
'exit' to end the conversation:
    >>>>>>>> NO HUMAN INPUT RECEIVED.
```